\documentclass{nature}

\usepackage{bm}
\usepackage{graphicx}

\usepackage{lineno}

\newcommand{\beeq}[1] {\begin{equation}#1\end{equation}}

\title{Intermittent dynamics in complex systems driven to depletion}

\author{Juan V Escobar$^{1,*}$ \& Isaac P\'erez Castillo$^{2,3}$}

\begin{document}

\maketitle

\begin{affiliations}
 \item Instituto de F\'isica, Universidad Nacional Aut\'onoma de M\'exico. Apdo. Postal 20-364, Cd. Mx., Mexico, C.P. 04510
 \item Departamento de F\'isica Cu\'antica y Fot\'onica, Instituto de F\'isica, Universidad Nacional Aut\'onoma de M\'exico. Apdo. Postal 20-364, Cd. Mx., Mexico, C.P. 04510
 \item London Mathematical Laboratory, 14 Buckingham Street, London WC2N 6DF, United Kingdom
 \item[*] Corresponding author: escobarjuanvalentin@gmail.com
\end{affiliations}

\begin{abstract}
When complex systems are driven to depletion by some external factor, their non-stationary dynamics can present an intermittent behaviour between relative tranquility and burst of activity whose consequences are often catastrophic. To understand and ultimately be able to predict such dynamics, we propose an underlying mechanism based on sharp thresholds of a local generalized energy density that naturally leads to negative feedback. We find a transition from a continuous regime to an intermittent one, in which avalanches can be predicted despite the stochastic nature of the process. This model may have applications in many natural and social complex systems where a rapid depletion of resources or generalized energy drives the dynamics. In particular, we show how this model accurately describes the time evolution and avalanches present in a real social system.
\end{abstract}

\section{Introduction}
Avalanches or bursts of activity are present in many natural and social systems that range from sand piles\cite{cite1}, earthquakes\cite{cite2}, solar flares\cite{cite3}, plastic deformation\cite{cite4}, domain flips in ferromagnets\cite{cite5}, species mass extinctions\cite{cite6}, neuronal activity \cite{cite7}, forest fires\cite{cite8}, landslides\cite{cite9}, clustering of self driven particles\cite{cite10} and hydrodynamics\cite{cite11}, to daily website hits\cite{cite12}, attendance to motion pictures\cite{cite13}, financial markets\cite{cite14} and book sales\cite{cite15}. Understanding and ultimately predicting the dynamics of these avalanches is of great scientific and practical importance, as they can lead to catastrophic events, of which large earthquakes and stock market crashes serve as two dramatic examples.

\begin{figure}
\begin{center}
\includegraphics[scale=1]{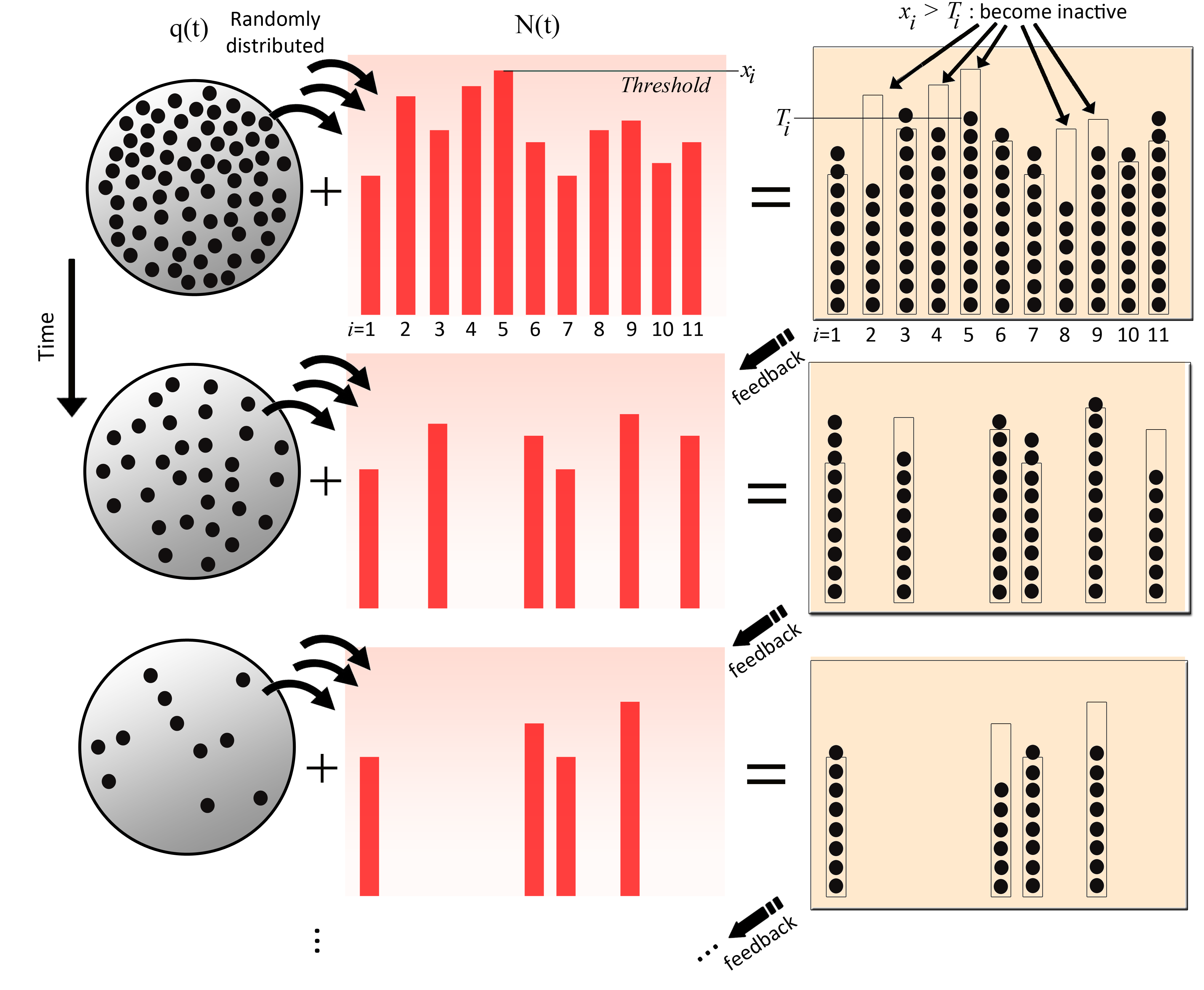} 
\caption{Schematic diagram of the negative feedback mechanism proposed:  A decaying input signal $q(t)$ is randomly distributed at time $t$ among $N(t)$ elements. When $T_i(t)$ crosses a threshold $x_i$, element $i$ becomes inactive, thus decreasing the overall number of active elements $N(t)$. To exemplify this mechanism, in the upper right panel elements labeled $i=2,4,5,8$ and $9$ become inactive at time $t+1$ since the condition $T_i(t)<x_i$ was fulfilled for each of these $i's$ at time $t$. }
\label{fig1}
\end{center}
\end{figure}

A very fruitful approach to understand the dynamics of many of these systems has been to assume local and relatively simple interactions of their constituting elements which  drive the system to a state that displays emergent behaviour. Importantly, this emergence results without the need of an external tuning parameter\cite{cite16,cite17}. This is the premise of  Self Organized Criticality (SOC)\cite{cite1}, for which models now abound. A  paradigmatic example is that of  earthquakes, which may be modelled as dissipative systems presenting SOC\cite{cite18, cite19}. There, a slow energy input eventually leads to relatively fast breakdown events spanning many orders of magnitude. Variations of this model\cite{cite20}, based on the Bak, Tang and Wiesenfeld (BTW)   sand-pile model\cite{cite21}, are able to recover the  main characteristics of earthquakes statistics. Roughly speaking, this class of BTW  models have two main ingredients. First, there is a continuous input of energy into the system that may continue indefinitely. Second, bursts of activity occur due to excess stress accumulation which is then toppled to neighbouring constituents giving rise to avalanches.

There exist, however, a plethora of other systems which are instead characterized by a steady decrease of their activity all the way to complete inertness. This happens as a result of a steady decay of the input energy, which translates into the system as an effective stress to which it must adapt, reaching at every step a new equilibrium configuration. While in some instances this is achieved in a relatively smooth way, in others the system is unable to adapt fast enough giving rise to large, often catastrophic events. Indeed, one may see an intermittent behaviour between relative tranquility and bursts of activity. Actually, it is now recognized that continuous driving forces can induce transitions between continuous and intermittent dynamics \cite{cite22}. This feature is of great importance, since systems that seem to be adapting well (smoothly) to such external stresses, may suddenly break down without any apparent warning signs. Examples of systems presenting this general behaviour range from the group of theaters showing a particular movie as weekly audience decreases, the number of surviving a mammalian genera during an extinction period or the price of a stock during a sell-off.

\section{Model definitions}

To model this behaviour we consider a system of non-interacting elements that receive (but do not accumulate) a share of the energy externally provided. Each element is pre-assigned a sharp threshold such that if the energy it has received is below this threshold, the element is permanently removed from the system. These thresholds are considered as intrinsic properties of the elements, and therefore do not change with time. We show below that this simple model naturally leads to a negative feedback responsible for the adaptation of the system as a whole. The general mechanism proposed is depicted in figure \ref{fig1}. This model reproduces both the smooth response as well as the intermittent one and provides a means of predicting catastrophic events in complex systems as they are driven to extinction. To be  mathematically precise, our model consists of $N$ basic elements labelled with $i=1,\ldots,N$. We then introduce a time-decaying external input signal $q(t)$, that, for sake of clarity, we assume takes integer values and decays rapidly with time. This function represents the total generalized available energy, which is randomly distributed among the $N(t)$ active elements at each time step $t$. An element belonging to this group becomes -and remains- inactive when the amount of energy allotted to it is below its own, pre-assigned threshold. To achieve this, a fixed set of time-independent thresholds $x_{i}>0$ for each of the constituent elements  $i=1,\ldots, N$ is drawn randomly from a distribution. We stress the point that at any given time step each element can be either active or inactive, but once it has become inactive it remains in that state. Let $\sigma_i(t)$ be the state of element $i$, with $\sigma_i(t)=0$ for the element being inactive, or one otherwise. If $N(t)=\sum_{i=1}^N\sigma_i(t)$ denotes the number of active elements at time $t$, the system evolves according to the following simple microscopic dynamical rule:
\beeq{
\sigma_i(t+1)=\sigma_i(t)\Theta\left(T_i(t)-x_i\right)\,.
\label{eq1}
} 
Here $\Theta(x)$ represents the Heaviside step function, and $T_i(t)$ is a portion of the input signal uniformly and randomly allotted only to active element $i$ such that $\sum_{i=1}^{N(t)} T_i(t)=q(t)$. Henceforth, we will call the set of variables $T_i(t)$ the occupancies. Notice that at a given time $t$, the probability of finding a particular configuration $\{T_1(t),\ldots, T_{N(t)}(t)\}$  follows a multinomial distribution with uniform probability $1/N(t)$. This implies that the average occupancy is the same for all elements and equal to  $T(t)\equiv{\rm E}[T_i]=q(t)/N(t)$, with  variance being  ${\rm Var}[T_i(t)]=q(t)(1-1/N(t))/N(t)$.  We take the initial conditions in which all constituents are active at $t=0$, that is $N(t=0)=N$, and define the activity for this model as the number of elements removed at each step $\Delta N(t)=N(t)-N(t+1)$.

\begin{figure}
\centering
\includegraphics[scale=1]{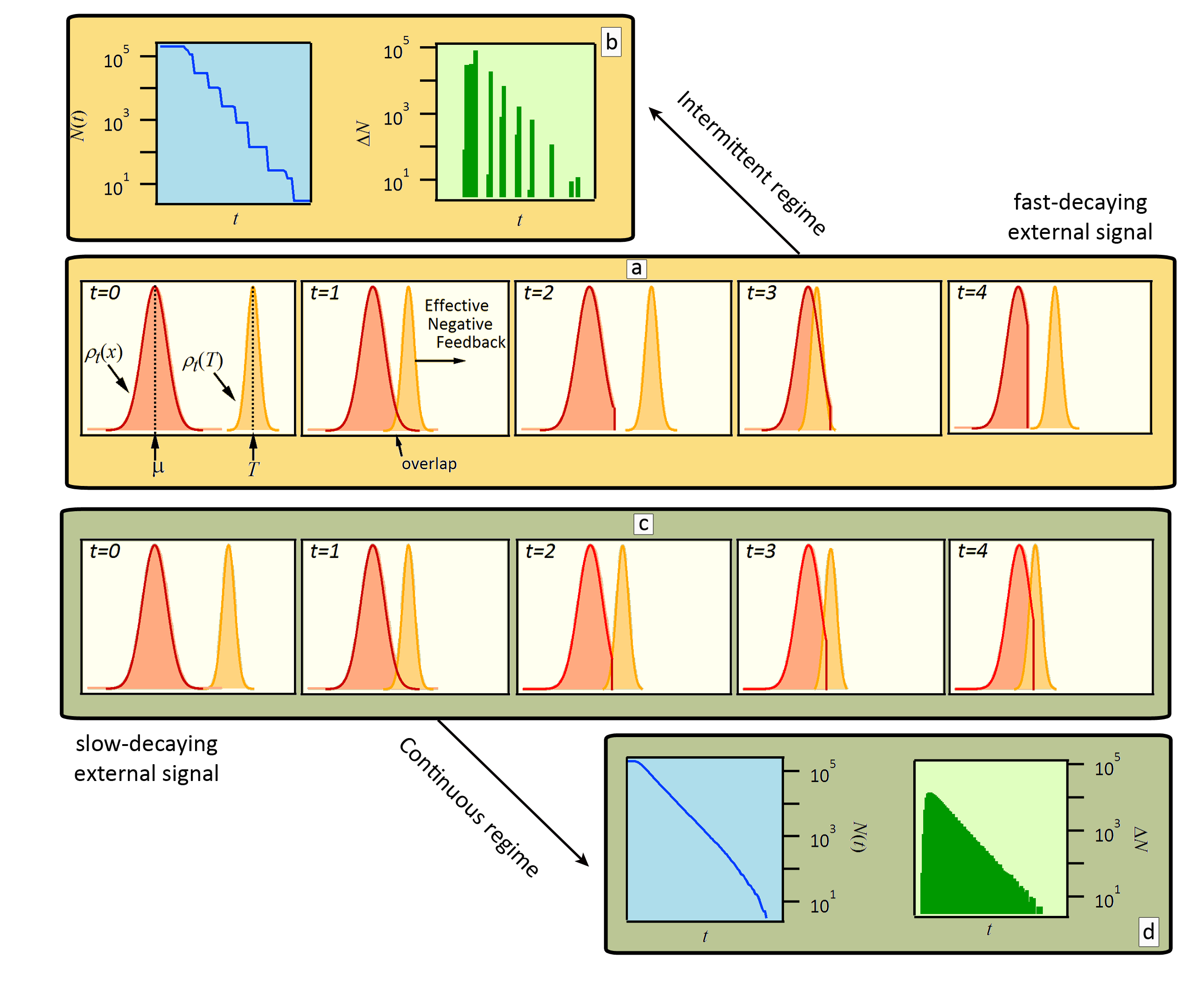}
\caption{Schematics of the time evolution of the number of active elements $N(t)$, represented by their distribution of thresholds $\rho_t(x)$ (solid line) and of the distribution of occupancies $\rho_t(T)$ (dashed line), evidencing the negative feedback that takes place in a) the intermittent and c) the continuous regimes. The dynamics is described in the main text. The corresponding typical activities, $\Delta N(t)$, for each regime are shown in figures b) and d). While in both examples the same $\rho_t(x)$ was chosen, it is the relatively faster decay rate of the input function what brings about intermittency in a) and b).}
\label{fig2}
\end{figure}

One can gain some intuition on how the system behaves under this mechanism by looking at the time evolution of the  density of active thresholds and the  density of occupancies, denoted as $\rho_t(x)$ and  $\rho_t(T)$ respectively. This will become useful to gain a good grasp on the mean-field solution to the model, presented below.  Figure \ref{fig2}a shows a sketch of the dynamical evolution of the system in terms of these two densities. Let us assume that at a given time these two distributions do not overlap (Figure \ref{fig2}a, $t=0$). Then, as time evolves, the distribution $\rho_t(T)$ moves to its left since the average occupancy $T(t)=q(t)/N(t)$ decreases over time, as $q(t)$ is a decreasing function and $N(t)$ has not changed yet. Eventually, the two distributions do overlap (Figure \ref{fig2}a, $t=1$). This means that  the overlapping fraction of active thresholds will become inactive, increasing the average occupancy $T(t)$, and pushing, in turn, the distribution $\rho_t(T)$ back towards higher values. As a consequence, a non-zero activity indirectly hinders any further activity, in what constitutes a negative feedback mechanism. Now again, the distributions are not overlapping (Figure \ref{fig2}a, $t=2$) but, as time evolves, this mechanism reoccurs (Figure \ref{fig2}a at $t=3$ and $t=4$). This sequence of events repeats itself until all elements have become inactive. Since in this sequence the system goes through periods of zero activity, Figure \ref{fig2}a actually illustrates its typical response when it is in the herein called intermittent regime. But with this representation it is also possible to foresee that if the external signal decays relatively slowly, the system will be able to adapt to the depletion of resources, so that small but finite portions of $\rho_t(x)$ are removed at every iteration (Figure \ref{fig2}c). We call this the continuous regime since the activity is always non-zero. The different behaviour of the system in both regimes can be appreciated in the plots of  $N(t)$ and $\Delta N(t)$ presented on Figures \ref{fig2}b and \ref{fig2}d for the intermittent and continuous regimes, respectively. These plots also help illustrate an important characteristic of the model: the maximum activity can be larger in the intermittent regime by over an order of magnitude as compared to the continuous one.

At this point it is important to contrast our model with the BTW model, as both present avalanches and are based on a sharp threshold dynamics as a response to some external energy input. In the BTW model, the elements $q_i(t)$ may represent the accumulated stress allocated in one of the $N(t)$ geological faults, and, as in our model, an earthquake or avalanche occurs when a fault crosses its threshold. But, unlike our model in which the elements remain inactive once they are brought to that state, as soon as a geological fault has toppled its excess stress, it can start again accumulating new stress. Another important difference is that in the BTW model the input signal keeps providing energy which the system continuously dissipates, while, in our case, the input signals evolves in times towards zero. Furthermore, our model does not allow any stress accumulation; in other words, it is history independent. Indeed, despite their shared characteristics, the actual mechanism giving raise to bursts of activity in each model is completely opposite in nature: in the BTW model avalanches result from a positive feedback mechanism, while in ours, a negative feedback one plays the central role.

\section{Results}

For concreteness, to quantify the behaviour of the model we  have chosen  an exponentially decaying external signal $q(t)=q_0 e^{-t/\tau}$ and proceeded to identify the continuous and intermittent regimes of the model in its space of parameters. As relevant parameters, apart of $\tau$, we have also chosen the initial coefficient of variation $\kappa\equiv \sigma/\mu$ of the distribution of active thresholds $\rho_{t=0}(x)$, where $\mu$ and $\sigma^2$ are its first two cumulants. We have subsequently estimated the dynamical phase diagram corresponding to the the microscopic dynamics of equation (\ref{eq1}) using two methods:  i) by deriving a simple dynamical mean-field equation and  ii) by Monte Carlo simulations.

\begin{figure}
\begin{center}
\includegraphics[scale=1]{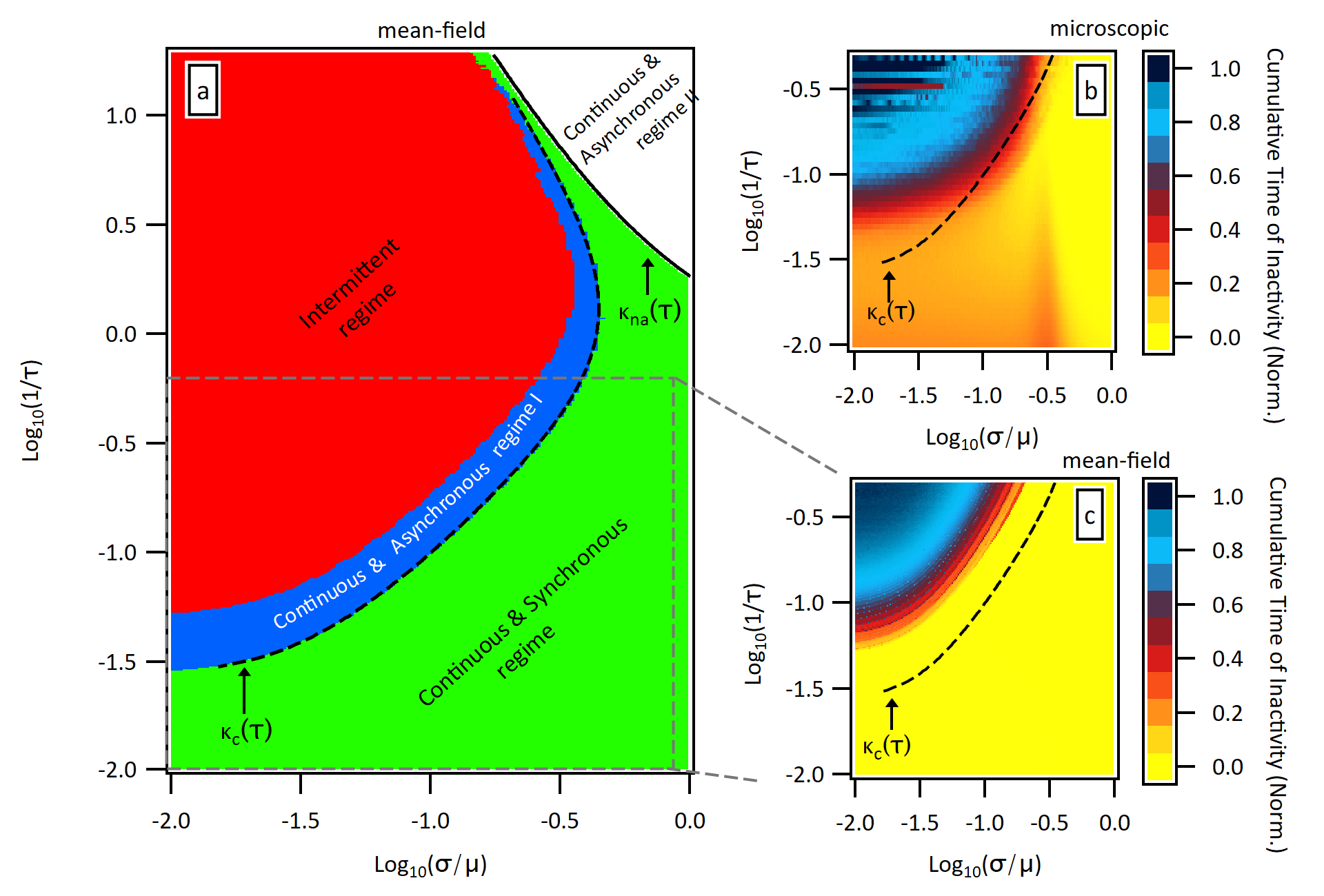} 
\caption{Analysis of dynamical phase-diagram. a) Phase diagram for the mean field model. Plots b) and c) show the intermittency phase diagram for the exact microscopicmodel and for the mean-field one, respectively. We also report the line $\kappa_{{\rm na}}(\tau)$ in figure a) and the line $\kappa_c(\tau)$ in all plots. The Monte Carlo simulations were carried out by taking $q(0)=2\times 10^{9}$, $N(0)=2\times10^{5}$, and $\mu=1000$, and subsequently varying the parameters $\tau$  and $\sigma$, to identify the regimes in  the phase diagram. Results of 50 different initial random distributions for $\rho_t(x)$ were averaged. For the mean field model, the system was considered to be inactive whenever $(N(t)-N(t+1))/N(t)<1\times10^{-5}$.}
\label{fig3}
\end{center}
\end{figure}

 Even though it is possible to obtain an exact effective macroscopic dynamics  using the method of the generating functional analysis a la Martin-Siggia-Rose\cite{cite0}, the resulting equations are rather cumbersome to analyse. Thus we have opted to derive a  simpler dynamical mean-field equation, which turns out to describe the exact microscopic dynamics remarkably well. After some algebra, a dynamical mean-field equation for the average number of active elements, which we also denote as  $N(t)$, is given by (see SI)
\beeq{
N(t+1)=\frac{N(t)}{2}{\rm erfc}\left(\frac{\mu-T(t)}{\sqrt{2(\sigma^2+T(t))}}\right)\,.
\label{eq:1}
}
To identify in which regime the model is  in the parameter space $(\kappa,1/\tau)$, we introduce two order parameters. The first one is a natural way to quantify intermittency through the normalized cumulative time of inactivity (see SI). In particular, this quantity is 0 when the system is in the continuous regime, and non-zero when in the intermittent one. The second parameter characterizes the ability of the system to follow the input signal and corresponds to see whether $z(t)\equiv \frac{\mu-T(t)}{\sqrt{2(\sigma^2+T(t))}}$ equals a constant $z_\star$ after some transient. By iterating the dynamical mean field equation (\ref{eq:1}) and following the values of these two parameters  we are able to identify four regimes or phases in the  $(\kappa,1/\tau)$-plane. Moreover, we are able to obtain analytical expressions for some of the transition lines. These four regimes are shown in Figure \ref{fig3}a. The intermittent regime, depicted in Figure \ref{fig3}a by a  red filled area, is characterised by the impossibility of the system to synchronise with the external signal, resulting in avalanches.  The continuous $\&$ asynchronous regime I (shown as the dark blue area in the same figure) corresponds to the system being again unable to synchronize to the external signal, but this time in a smooth way, \textit{i.e.},  avalanches are no longer generated. On the other side, in the continuous $\&$ synchronous regime, shown by the green area in Figure \ref{fig3}a, the system is able to follow the external signal. Finally, the continuous $\&$ synchronous regime II, captures the impossibility of the system to match the external signal due solely to restrictions in the initial conditions $T(t)>0$.

Notice that, within the mean-field approximation, it is possible to obtain analytically the bifurcation line, denoted as $\kappa_c(\tau)$, separating the asynchronous to the synchronous regimes, as well as the bifurcation line $\kappa_{{\rm na}}(\tau)$ separating the synchronous to the asynchronous regime II. The first line is derived by assuming that the synchronous regime is characterised by $z(t)=z_\star$ being a constant. This implies, in turn, that $z_\star$ and $\tau$ are related by the formula $z_\star(\tau)={\rm erfc}^{-1}(2 e^{-1/\tau})$. 

Naturally, the transition to the intermittent regime is derived by assuming a continuous bifurcation of $z(t)$ around $z_{\star}$.  After some algebra, the  line  separating both regimes in the $(\kappa,1/\tau)$-plane is given by the following set of coupled equations:
\begin{eqnarray}
\left\{\begin{array}{l}
e^{z_\star^2}{\rm erfc}\left(z_\star\right)=\frac{(\mu+2(\kappa\mu)^2+T_\star)T_\star}{2\sqrt{2\pi}((\kappa\mu)^2+T_\star)^{3/2}}\\
z_\star(\tau)=\frac{\mu-T_\star}{\sqrt{2((\kappa\mu)^2+T_\star)}}
\end{array}\right.\,,
\label{eq:bc}
\end{eqnarray}
  with $\frac{1}{\tau}=-\log\left[\frac{{\rm erfc}(z_\star)}{2}\right]$. Thus, given $\mu$,  the solution of (\ref{eq:bc}) results in a pair $(\kappa_{c}(\tau),T_\star(\tau))$, where the first function $\kappa_c(\tau)$ corresponds to the bifurcation line separating the aforementioned regimes, while line $T_\star(\tau)$ tells which value $T(t)$ takes precisely at the transition. As shown in Figure \ref{fig3}a, the agreement between the bifurcation line (\ref{eq:bc}) and the dynamical phase diagram obtained from the same equation is excellent (see the SI for more details on the synchronization plots for this and other values of $\mu$). The second bifurcation line, as reported in Figure \ref{fig3}a,  follows from the fact that in the continuous  and synchronous regime we must have $z(t)=z_\star=[\mu-T_\star]/\sqrt{2(\sigma^2+T_\star)}$ and, since $N(t)$ decays precisely like the signal $q(t)$, one can easily derive from equation (\ref{eq:1}) that $1/\tau=-\log\left[\frac{1}{2}{\rm erfc}\left(\frac{\mu-T_\star}{\sqrt{2(\sigma^2+T_\star)}}\right)\right]$. However, as $T_\star\geq 0$ there is a part of the phase diagram inaccessible in the synchronous regime,  given by the following line $\kappa_{{\rm na}}(\tau)$ in the $(\kappa, 1/\tau)$-plane (see SI):
\beeq{
\frac{1}{\tau}=-\log\left[\frac{1}{2}{\rm erfc}\left(\frac{1}{\sqrt{2}\kappa_{\rm na}}\right)\right]
}

 How well the mean field description captures the microscopic model can be appreciated by comparing Figures \ref{fig3}b and \ref{fig3}c, where we have calculated the cumulative time of inactivity for each model, using Monte Carlo simulations for the microscopic one. Despite the simplicity of equation (\ref{eq:1}), the agreement with the microscopic model is reasonably good. Notice that these plots show  there is a clear border separating the intermittent and synchronous regimes, and that, in general, intermittency can be avoided for faster decays with relatively wide $\rho_t(x)$ distributions, as expected.  Furthermore, both models present a gap: below a critical value of $1/\tau$, the system is continuous regardless of how small $\kappa$ is. The magnitude of this gap is actually a function of $\mu$, and its analytical expression can be found for the mean-field description (see SI). We point out that, unlike the mean-field model, the microscopic one does not present a synchronized regime. Rather, in the latter the activity always lags behind the external signal, \textit{i.e.}, it is always marginally desynchronized. We find, however, that the activity does become less synchronized in the intermittent regime, but no clear transition is observed for this model.

\begin{figure}
\begin{center}
\includegraphics[scale=1]{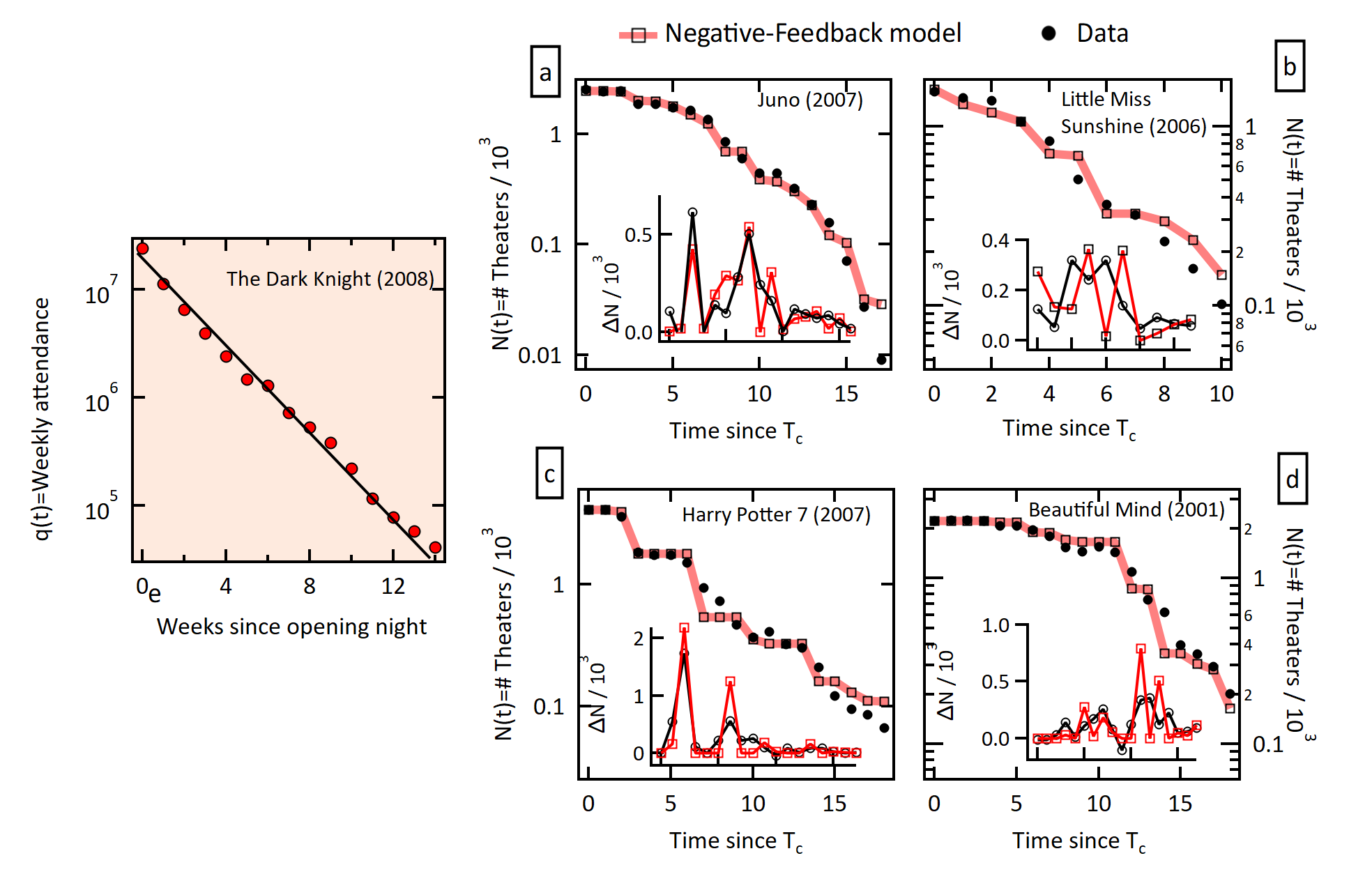} 
\caption{Application of the model to real data. Left plot: Attendance to the movie The Dark Knight (2008). The initial attendance is $q(0)$=\$238,615,211/\$7.18, where  \$7.18 is the average ticket price across the U.S. in 2008, evidencing an exponential decay nature of $q(t)$ in this system. Right panel: Negative feedback avalanche examples. Actual (dots) and calculated (squares and thick line) number of theaters vs. time for four different movies. The corresponding insets illustrate the time evolution of the avalanche magnitudes (or activity) $\Delta N$. The parameters used are fixed for all movies ($\mu=400$, $\sigma=0.45\mu$), while the average ticket prices employed are \$5.66, \$6.88 and \$7.18 for the years 2001, 2006 and 2007 respectively. $T_c$ is the time in weeks since the maximum attendance was registered, which does not necessarily coincide with the opening week. All data was obtained from www.boxofficemojo.com. See Ref.\cite{cite13} for details.}
\label{fig4}
\end{center}
\end{figure}

\section{Application to a real system}

Despite its apparent simplicity, the model presented here is robust enough to be applicable to real complex systems. As a proof of principle, we now implement it to investigate the dynamics of the supply of theaters for a given movie that played in the US from 1970 to 2000. The dynamics of the attendance to movie theatres in the U.S. has been studied recently in Ref.\cite{cite13}, where it was shown that the attendance is in general very well described by a decaying exponential (see example of left plot in Figure \ref{fig4}) as inferred from the weekly revenue of a movie (see SI), and is thus ideally suited to test our model. This system meets our model if we consider that as $q(t)$ people fill up the $N(t)$ theaters at time $t$, a theater removes the movie at time $t+1$ if the local attendance $T_i(t)$ is lower than its threshold $x_i$. In other words, a theater removes a movie after the revenue from one week is below some pre-established value. We assume now that the distribution of local thresholds is again normally distributed. As a first order approximation we assume that at time $t$ the whole available population $q(t)$ is randomly distributed over the available theaters $N$. We realize that this is an enormous simplification of the actual problem since, of course, geographic restrictions limit the available theaters for people. Nevertheless, similar results as the ones presented below are obtained by sectioning the data and assigning smaller variance per section.

The right panel in Figure \ref{fig4} presents four examples of how our model (thick pink line and square symbols) is capable of reproducing the theater dynamics (black dots) remarkably well, in some cases spanning almost three orders of magnitude. Note the logarithmic scale on the number of theater axis of this figure. Both small and relatively large events are equally reproducible. In all examples shown, the same parameters $\mu=400$, $\sigma=0.45\mu$, were used. In other words, we only require the initial number of theaters to reproduce the whole time series of $N(t)$ using as input the specific time series for $q(t)$ and a normal distribution of thresholds. Other than the fixed choice of distribution of thresholds, there are no free parameters in this model.

\begin{figure}
\begin{center}
\includegraphics[scale=1]{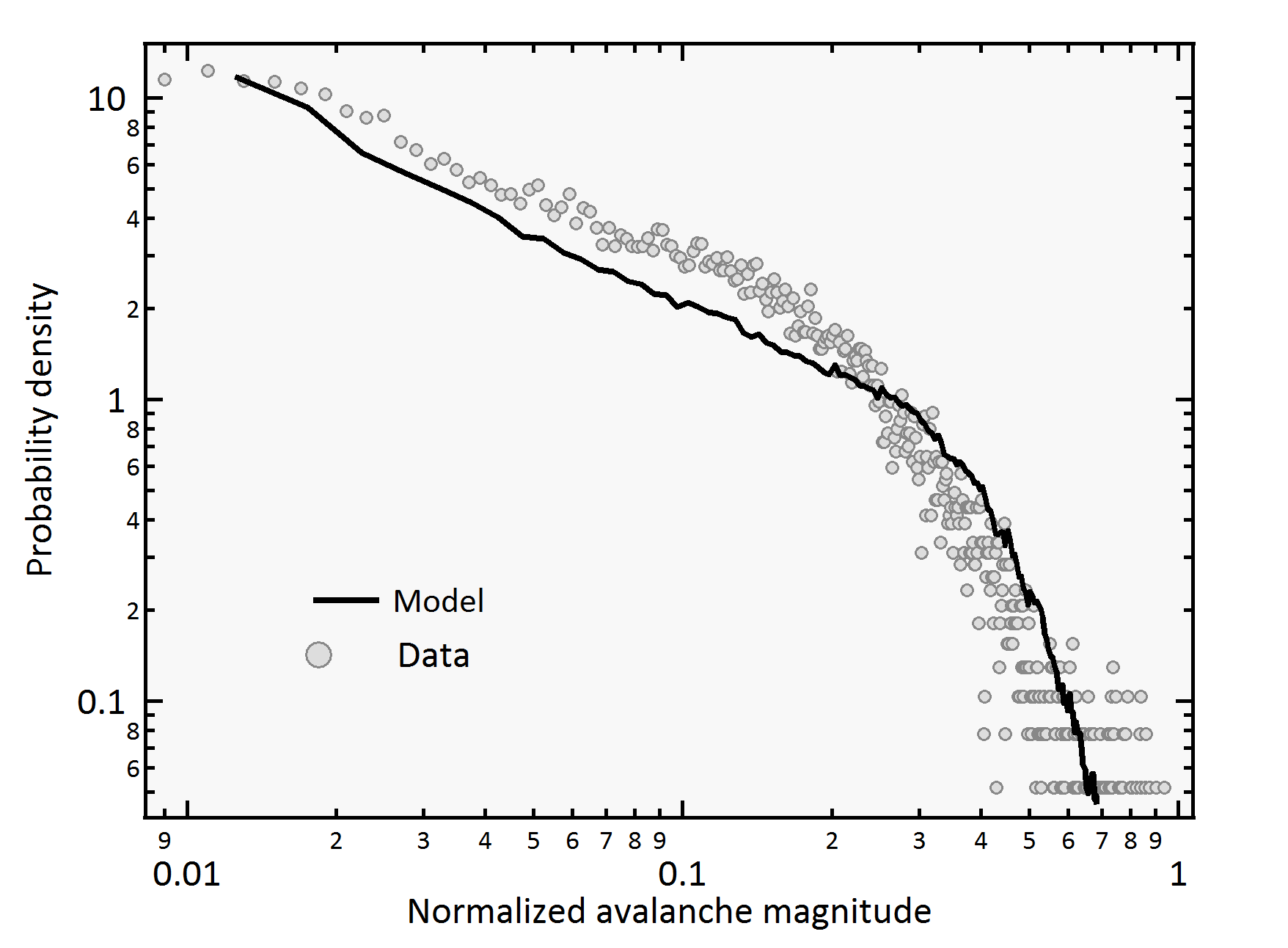} 
\caption{Avalanches in this complex system. Probability density of the avalanche magnitude for the actual data (circles) and for the reproduced data (line) obtained with the microscopic model. Avalanche magnitudes were normalized by the corresponding $N(0)$. A total of 100 runs per movie were averaged to cancel out any special condition of the initial threshold distributions. The distributions $\rho_t(x)$ were chosen randomly from a normal distribution with fixed ($\mu=400$, $\sigma=0.45$) for all movies. Note that the model reproduces both small and large events.}
\label{fig5}
\end{center}
\end{figure}

We now apply the model to a collection of 3000 different movies that played in the U.S from 1970 to 2000 (see SI). We postulate that this social system is characterized by a single distribution of thresholds, and therefore, the same parameters $\mu=400$, $\sigma=0.45\mu$ are used in all instances, and only $q(t)$ and $N(0)$ are different for each movie and determined by external factors (see SI for further details). To assess the capability of the model to completely recover the dynamics of this social system we obtain the distribution of $\Delta N(t)$ normalized by $N(0)$ for each movie from both the model (solid line in figure \ref{fig5}) and the actual data (dots in figure \ref{fig5}). Notice that the model recovers the whole shape of the distribution of these normalized avalanches. This distribution is actually well fit to a power law with an exponential tail (not shown).
In this particular system, an epidemic branching mechanism is responsible for the exponential decay of the effective energy $q(t)$, Ref.\cite{cite13}. It is possible that other complex systems also present a similar rapid decline of resources, driving the dynamics towards intermittency.

\section{Conclusions}

To conclude, we have introduced a negative feedback model with sharp thresholds to describe the dynamics of complex systems driven to depletion, which is simple enough to be amenable to analytical treatment and yet powerful enough to accurately predict the dynamics in real complex systems. We foresee that this model may be used to explain and predict small and large avalanches in other systems for which there is an exponentially fast decline of some quantity playing the role of an energy source randomly distributed among elements that need a certain minimum amount of it to remain active (or survive). Since large avalanches are indeed catastrophic in many social and natural settings, implementing variations of the model presented in this work could help prevent them.

\begin{addendum}
\item[Supplementary Information] is linked to the online version of the paper at www.nature.com/nature
\item J. V. E. is grateful to Denis Boyer and Leonardo Dagdug for interesting discussions about the model. J. V. E gratefully acknowledges funds from DGAPA-UNAM No. IA101216 and  No. IA103018, as well as from CONACYT No. CB 2013/221235. 
\item[Author Contributions] J. V. E. proposed the model.  I.P.C. solved the model exactly and helped with the analytical understanding. Data and numerical analysis was done by J. V. E. The authors discussed the results and commented the manuscript.
 \item[Competing Interests] The authors declare that they have no competing interests.
 \item[Correspondence] Correspondence and requests for materials should be addressed to Juan V. Escobar ~(email: escobar@fisica.unam.mx).
\end{addendum}

\bibliography{refsDraft}

\end{document}